\documentclass[aps,prl,preprint,groupedaddress]{revtex4-1}
\usepackage{graphicx}
\usepackage{amsmath}
\usepackage{setspace}
\usepackage{color}
\usepackage{soul}

\begin{document}
\title{Frictionless Zonal Flow Saturation by Vorticity Mixing}
\author{J. C. Li}
\affiliation{CASS, University of California, San Diego, California 92093, USA}

\author{P. H. Diamond}
\affiliation{CASS, University of California, San Diego, California 92093, USA}

\begin{abstract}	
	Consideration of wave--flow resonance addresses the long-standing problem of how zonal flows (ZF) saturate in the limit of weak or zero frictional drag, and also determines the ZF scale. 
	For relevant magnetic geometries, the frequently quoted tertiary instability requires unphysical enhancement of ZF shear and thus is irrelevant to the near-marginal, frictionless regime.
	We show that resonant vorticity mixing, which conserves potential enstrophy, enables ZF saturation in the absence of drag, and so is effective in the Dimits up-shift regime. 
	Vorticity mixing is incorporated as a nonlinear, self-regulation effect in an \textit{extended} 0D predator--prey model of drift--ZF turbulence.
	This analysis determines the saturated ZF shear and shows that the mesoscopic ZF width scales as
	$L_{ZF}\sim f^{3/16} (1-f)^{1/8} \rho_s^{5/8} l_0^{3/8}$ 
	in the relevant adiabatic limit (i.e., $\tau_{ck} k_\|^2 D_\| \gg 1$).
	$f$ is the fraction of turbulence energy coupled to ZF and $l_0$ is the mixing length absent ZF shears.
	We calculate and compare the stationary flow and turbulence level in frictionless, weakly frictional, and strongly frictional regimes. 
	In the frictionless limit, the results differ significantly from conventionally quoted scalings derived for frictional regimes.
	The flow is independent of turbulence intensity. 
	The turbulence level scales as $E \sim (\gamma_L/\varepsilon_c)^2$, 
	which defines the extent of the ``near-marginal" regime to be $\gamma_L < \varepsilon_c$, for the case of avalanche-induced profile variability. 
	Here, $\varepsilon_c$ is the rate of dissipation of potential enstrophy and $\gamma_L$ is the characteristic linear growth rate of fluctuations.
	The implications for dynamics near marginality of the strong scaling of saturated $E$ with $\gamma_L$ are discussed.
\end{abstract}
\maketitle

Predicting turbulence and transport in states evolving from saturated instability is the goal in many areas of research on nonlinear dynamics. 
Examples include pipe flow and the question of friction factor scaling\cite{Moody_1944}, Rayleigh--B{\'e}nard convection and the question of the scaling of Nusselt number with Rayleigh and Prandtl numbers\cite{Berge_1984}, and drift wave turbulence in confined plasmas\cite{Garbet_PPCF_2004} and the question of turbulent diffusivity scaling\cite{McKee_NF_2001} with $\rho_* \equiv \rho_0/a$. 
A frequently operative saturation mechanism works by the coupling of free energy to stable secondary patterns by the interaction of primary instabilities. 
The secondary patterns form as a result of secondary instability. 
When the structure of the secondary patterns drives the tertiary instability above threshold, the onset of such tertiary modes causes further relaxation.
Zonal flow (ZF) generation in quasi-geostrophic (QG)\cite{Charney_1948} and drift wave (DW)\cite{Diamond_PPCF_2005} turbulence is a prime example of saturation-by-secondary. 
Indeed, zonal flows are very effective at regulating DW turbulence, as they are the secondary modes of minimal inertia, transport, and damping\cite{Diamond_PPCF_2005}. 
Such a mechanism naturally can be thought of as a predator--prey type ecology\cite{Diamond_PRL_1994,Kobayashi_PoP_2015}, in which the secondary `predator' feeds of (i.e., extracts energy from) the primary `prey'.
In such a system, the damping of the predator (here, the ZF) ultimately regulates the full system. 
Drag, due to bottom friction in QG or because of collisions in plasmas, is usually invoked to damp ZF. 
However, this picture is unsatisfactory for present day and future regimes of low collisionality. 
Thus, it becomes essential to understand \textit{frictionless} ZF saturation and its implications for DW turbulence. 
Of course, ZF saturation significantly impacts transport and turbulence scalings.
Note that understanding scalings in the frictionless regime is essential for developing reduced models thereof. 
Taken together, these are the topics of this paper. 

We study drift--ZF turbulence with special focus on the frictionless regime where the flow drag approaches zero. 
Note that the DW drive--which can depend on electron collisionality--is not affected by the distinction between frictional and frictionless ion regimes, since frictional damping of drift \textit{waves} is weak.
Many works on ZF generation\cite{Diamond_PPCF_2005,Gurcan_ZF_2015,Staebler_PoP_2004} exist, but the question of how ZF saturates, absent frictional drag, remains open.
\textit{We show that turbulent mixing of zonal vorticity by drift waves in the presence of ZF saturates secondary flows for near-marginal turbulence} (with low to zero frictional drag), \textit{and thus is effective at regulating the Dimits up-shift regime. }
The Dimits regime\cite{Dimits_PRL_1996} is that of a frictionless DW--ZF system close to the linear instability threshold, where nearly all the energy of the system is coupled to ZF, so that the residual transport and turbulence are weak, though finite.
This induces an up-shift in the onset of the turbulent fluxes when plotted vs $\nabla T$.
Turbulent vorticity mixing is fundamentally different from viscous flow damping.
Turbulent vorticity mixing conserves total potential enstrophy (PE) between the mean field--i.e., the zonal component--and fluctuations. 
In contrast, the flow viscosity dissipates both the ZF and (DW flow) fluctuations.
Note that even though there exist many simulation studies on the DW--ZF dynamics\cite{Ernst_PoP_2009,Lang_PoP_2008,Wang_PoP_2006,Jenko_PoP_2002}, few of them address the question of how ZFs and thus the system saturate in the frictionless limit for near-marginal flux driven turbulence.

Though sometimes mentioned in this context, tertiary instability is \textit{not} effective for most cases of ZF saturation as it is strongly suppressed by magnetic shear. 
Indeed, in simulation studies, onset of tertiary instability requires an \textit{artificial} increase in the ZF shearing rate\cite{Rogers_PRL_2000} so as to overcome magnetic shear.
Ion temperature gradients can provide an extra source of free energy to drive the tertiary mode, in addition to flow shears.
However, such a contribution to the growth rate of the tertiary mode is of order $O(k^2 \rho_i^2)$, and thus does not qualitatively alter tertiary stability\cite{Kim_PoP_2002}.
Tertiary instability of ZF may occur in flat-q regimes\cite{Mantica_PRL_2011} with weak magnetic shear. 
Even there, the onset of tertiary mode of ZF, i.e., the Kelvin--Helmholtz instability, requires the ZF shear to exceed a threshold\cite{Chiueh_PoF_1986}. 
However, in the cases we have encountered, the ZF shear is always below the threshold. 
Moreover, we are interested in the Dimits up-shift regime, where the tertiary mode is stable.
Therefore, in this work, we shift the paradigm from the hypothetical saturation induced by tertiary instability to saturation by vorticity mixing.

Turbulent vorticity mixing is driven by DW--ZF resonance. 
The resonance response is characterized by this auto-correlation time due to the resonant mixing, i.e., $\delta (\omega_k - k_y \langle v_y \rangle) \sim \tau_{ck}$.
It is analogous to Landau damping-induced absorption of plasmons during collapse of Langmuir turbulence\cite{Galeev_1977,Che_PNAS_2017}. 
This analogy is developed in the supplemental material\cite{supplement}.

As a simple example, we study the Hasegawa--Wakatani DW system\cite{Hasegawa_PRL_1983} in slab geometry with ZF $\langle v_y\rangle$: 
\begin{equation} \label{eq:density}
	\left( \frac{d}{dt} + \tilde{\bf{v}}_E \cdot \nabla \right ) \tilde{n} 
	+ \tilde{v}_x \frac{\nabla n_0}{n_0}
	= D_\parallel \nabla_\parallel^2 (\tilde{n}-\tilde{\phi})
	+ D_c \nabla^2 \tilde{n},
\end{equation}
\begin{equation} \label{eq:vorticity}
	\left( \frac{d}{dt} + \tilde{\bf{v}}_E \cdot \nabla \right ) \tilde{\rho}
	+ \tilde{v}_x \langle \rho \rangle '
	= D_\parallel \nabla_\parallel^2 (\tilde{n}-\tilde{\phi})
	+ \chi_c \nabla^2 \tilde{\rho},
\end{equation}
where we define $D_\parallel \equiv v_{The}^2/\nu_{ei}$ and
$d/dt \equiv \partial_t + \langle v_y\rangle \partial_y$. 
Here, $\langle \cdot \rangle$ is zonal average, i.e., $\langle \cdot \rangle \equiv \int dy /L_y$. 
$\nu_{ei}$ is electron--ion collision frequency and $v_{The}$ is electron thermal speed.
We have normalized electric potential fluctuation as $\tilde{\phi} \equiv e \delta \phi / T_e$ and density fluctuation as $\tilde{n} \equiv \delta n / n_0$, where $n_0$ is the equilibrium density.
The magnetic field is uniform and lies in $\hat{z}$ direction. Both $n_0$ and $\langle v_y \rangle$ vary only in $\hat{x}$ direction.
$\tilde{\rho}\equiv \rho_s^2 \nabla_\perp^2 \tilde{\phi}$ is the vorticity fluctuation, where $\rho_s$ is the ion Larmor radius at electron temperature,
$\langle \rho \rangle \equiv \langle v_y\rangle' \rho_s / c_s$ is the zonal vorticity where $c_s$ is the ion sound speed.
$\tilde{\bf{v}}_E \equiv c_s \hat{\bf{z}} \times \nabla \tilde{\phi}$ 
is the $E\times B$ velocity fluctuation.
$D_c$ and $\chi_c$ are the collisional particle diffusivity and vorticity diffusivity (i.e., viscosity).
The DW is the dominant instability population, because the vorticity gradient drive is quantitatively weaker than the $\nabla n_0$ drive, i.e., 
$k_y \rho_s^2 \langle v_y\rangle''/\omega_{*e} \ll 1$ where $\omega_{*e}\equiv k_y \rho_s c_s / L_n$ is the electron drift frequency and $L_n\equiv n_0 / |dn_0/dx|$ is the density gradient scale.
The Hasegawa--Wakatani model is simple but generic.
It is a \textit{general} model of electron DW turbulence. 
Its structure is similar to \textit{any} electron DW system, such as collisional DW, dissipative and collisionless trapped electron modes.
The tertiary mode discussed here is driven by the zonal flow shear. Therefore, its free energy source has been taken into account in the Hasegawa--Wakatani drift wave model studied in this work.

The generation and saturation of ZF by drift waves are described by PV (potential vorticity) mixing. 
The fluctuating PV is defined as $\tilde{q}\equiv \tilde{n}-\tilde{\rho}$, and the zonal PV is $\langle q\rangle \equiv \langle n \rangle -\langle \rho \rangle$.
Hence, the evolution equation for fluctuating PV can be obtained by subtracting Eq. \eqref{eq:vorticity} from Eq. \eqref{eq:density}, yielding
\begin{equation} \label{eq:PV_fluc}
	\left( \frac{d}{dt} + \tilde{\bf{v}}_E \cdot \nabla \right) \tilde{q} 
	+ \tilde{v}_x \frac{\partial}{\partial x} \langle q \rangle
	= D_{q,c} \nabla^2 \tilde{q}.
\end{equation}
Here, $D_{q,c} \sim (D_c + \chi_c)/2$ is the collisional diffusivity of PV.
In multiplying both sides of Eq. \eqref{eq:PV_fluc} by $\tilde{q}$, we obtain the PE equation\cite{Ashourvan_PRE_2016,Ashourvan_PoP_2017}:
\begin{equation}
	\frac{\partial}{\partial t} \Omega
	= - \frac{\partial}{\partial x} \frac{\langle \tilde{v}_x \tilde{q}^2 \rangle}{2}
	- \langle \tilde{v}_x \tilde{q} \rangle \frac{\partial}{\partial x} \langle q \rangle
	- \Gamma_{q}^{\text{Ter}} \frac{\partial}{\partial x} \langle q \rangle
	- \varepsilon_c \Omega^{3/2}
	+ \gamma_L \Omega,
\end{equation}
where $\Omega \equiv \langle \tilde{q}^2 \rangle/2$.
$\Gamma_{q}^{\text{Ter}} = - D_{NL} \partial_x \langle q \rangle$ is the PV flux driven by tertiary instability.
$D_{NL}$ is the PV diffusivity caused by tertiary modes and depends on $\langle \rho \rangle$.
The turbulent PE flux is due to nonlinear spreading, and can be approximated as a diffusive flux, i.e., 
$\langle \tilde{v}_x \tilde{q}^2 \rangle/2 \sim - D_\Omega \partial_x \Omega$\cite{Ashourvan_PRE_2016}.
The nonlinear PE dissipation $\varepsilon_c \Omega^{3/2}$ represents the forward cascade (to dissipation) of PE. 
$\epsilon_c$ is the nonlinear dissipation rate of potential enstrophy.
$\gamma_L$ is the characteristic linear growth rate of DW, which drives the turbulence and thus produces PE. 
The coupling of PV flux and zonal PV profile gradient conserves PE between mean field and fluctuations.

The equations for mean-field density and zonal vorticity are
\begin{equation}
	\frac{\partial}{\partial t} \langle n \rangle = 
	- \frac{\partial}{\partial x} \langle \tilde{v}_x \tilde{n} \rangle
	+  \frac{\partial}{\partial x} D_{NL}  \frac{\partial}{\partial x} \langle n \rangle
	+ D_{c} \nabla^2 \langle n \rangle,
\end{equation}
\begin{equation}
	\frac{\partial}{\partial t} \langle \rho \rangle = 
	- \frac{\partial}{\partial x} \langle \tilde{v}_x \tilde{\rho} \rangle
	- \mu_{c} \langle \rho \rangle
	+  \frac{\partial}{\partial x} D_{NL}  \frac{\partial}{\partial x} \langle \rho \rangle
	+ \chi_{c} \nabla^2 \langle \rho \rangle.
\end{equation}
$\mu_c$ is frictional drag coefficient.
Onset of tertiary instability can be included in reduced models, if needed. However, here we neglect it, because the relevance of such tertiary modes to ZF saturation in confinement devices is debatable.

To close the system, we need to calculate the turbulence-driven fluxes. 
Detailed derivations of PV and particle fluxes can be found in Ref. \cite{Ashourvan_PRE_2016}.
The frictionless zonal flow saturation is directly determined by the turbulent diffusion of vorticity. In order to calculate the vorticity flux, we first calculate the PV flux, since PV is conserved by the Hasegawa--Wakatani drift wave system discussed here.
The quasilinear PV flux is diffusive, i.e., 
$\langle \tilde{v}_x \tilde{q} \rangle 
	= - D_{q,turb} \partial_x \langle q \rangle$,
which is obtained from Eq. \eqref{eq:PV_fluc}, neglecting collisional diffusion.
Ref. \cite{Ashourvan_PRE_2016} treats the turbulent diffusion of vorticity as non-resonant.
Here, the turbulent diffusivity of PV has a resonant part and a non-resonant part, i.e., 
$D_{q,turb} = D_{q}^{\text{res}} + D_{q}^{\text{non-res}}$.

When the flow velocity of zonal flows approaches the phase velocity of drift waves, the turbulent diffusion of PV is resonant.
The resonant response is determined by the propagator $\delta (\omega_k - k_y \langle v_y \rangle)$, where $\omega_k$ is the drift wave frequency.
The resonant diffusivity of PV is set by the resonance between DW phase velocity and the local ZF profile, which yields
$ D_q^{\text{res}} = \sum_k |\tilde{v}_{x,k}|^2 \pi \delta(\omega_k - k_y \langle v_y \rangle)$,
where $\tilde{v}_{x,k}$ is the fluctuating velocity in the radial direction and $\omega_k$ is the DW frequency. 
The resonant scattering here has a characteristic spectral autocorrelation time scale
$
\tau_{ck} \sim |\Delta(\omega_k - k_y \langle v_y \rangle)|^{-1}
\sim \left\{ |(v_{g,y} - v_{ph,y}) \Delta k_y |
+|v_{g,x}\Delta k_x|\right\}^{-1}
$, 
where we have used $\langle v_y \rangle \cong \omega_k / k_y = v_{ph,y}.$
Here, $\Delta(\cdot)$ is an operator determining the width of turbulence spectrum when plotted against the content of the operator.
The resonance is between drift waves and the instantaneous ZF profile. Thus, this autocorrelation time is shorter than the time scale of ZF evolution, i.e., $\tau_{ck} \ll \tau_{ZF}$, consistent with ZF evolution by turbulent PV mixing.
The correlation time $\tau_{ck}$ is shorter as compared to the 1D case, where the spectral width is due to the mismatch between group velocity and phase velocity, i.e., $\tau_{ck} \sim |(v_{g} - v_{ph}) \Delta k |^{-1}$, only. 
As a result, the resonant diffusivity is
$
D_q^{\text{res}} = \sum_k k_y^2 \rho_s^2 c_s^2|\phi_k|^2 \tau_{ck}.
$

The non-resonant diffusivity can be obtained by quasilinear theory, and is
$
D_q^{\text{non-res}} = \sum_{\omega_k \neq k_y \langle v_y \rangle} k_y^2 \rho_s^2 c_s^2|\phi_k|^2
|\gamma_k|/|\omega_k - k_y \langle v_y \rangle|^2
$.
$\gamma_k$ is the linear growth rate of DW. 
The Doppler shifted frequency of the DW is
$
\omega_k
\cong \omega_{*e}/(1+k_y^2\rho_s^2+L_m^{-2}\rho_s^2)
$
and the growth rate is
$
\gamma_k \cong (\omega_{*e}^2/k_\parallel^2 D_\parallel) (k_y^2\rho_s^2+L_m^{-2}\rho_s^2)/(1+k_y^2\rho_s^2+L_m^{-2}\rho_s^2)^3
$.
Both of them depend upon the mode structure. 
$L_m\equiv \langle k_x^2 \rangle ^{-1/2} $ 
is the eigenmode scale in radial direction.
Thus, the non-resonant diffusivity depends on the mode scale, which yields
\begin{equation}
	D_q^{\text{non-res}} \cong 
	\sum_{k \text{ for which } \omega_k \neq k_y \langle v_y \rangle} 
	\frac{k_y^2 \rho_s^2 c_s^2}{ k_\parallel^2 D_\parallel}
	\frac{k_y^2 \rho_s^2 + L_m^{-2}\rho_s^2}{1+k_y^2 \rho_s^2 + L_m^{-2}\rho_s^2}
	|\phi_k|^2.
\end{equation}
The non-resonant diffusivity is negligible in comparison to the resonant diffusivity.
The ratio of the two is $D_q^{\text{non-res}}/D_q^{\text{res}} \sim (k_\parallel^2 D_\parallel \tau_{ck})^{-1} (k_y^2 \rho_s^2 + L_m^{-2}\rho_s^2)/(1+k_y^2 \rho_s^2 + L_m^{-2}\rho_s^2) < (k_\parallel^2 D_\parallel \tau_{ck})^{-1}$.
In the relevant regime where $k_\parallel^2 D_\parallel \tau_{ck} \gg 1$, we obtain $D_q^{\text{non-res}} \ll D_q^{\text{res}}$, i.e., the mixing of PV is primarily \textit{resonant}.

The turbulent particle flux driven by DW turbulence in the adiabatic regime is diffusive, i.e., 
$ \langle \tilde{v}_x \tilde{n} \rangle
	= - D_{n,\text{turb}} \partial_x \langle n \rangle$,
where 
\begin{equation}
	D_{n,\text{turb}} = \sum_k
	\frac{k_y^2 \rho_s^2 c_s^2}{ k_\parallel^2 D_\parallel}
	\frac{k_y^2 \rho_s^2 + L_m^{-2}\rho_s^2}{1+k_y^2 \rho_s^2 + L_m^{-2}\rho_s^2}
	|\phi_k|^2.
\end{equation}
The key propagator in the particle flux is $1/(\omega_k - k_y \langle v_y \rangle + i k_\parallel^2 D_\parallel)$. In the relevant regime of near-adiabatic electrons, i.e., $k_\parallel^2 D_\parallel \gg \omega_k$, this propagator is determined by $1/ i k_\parallel^2 D_\parallel$ and does not involve the resonance.
As a result, the resonance does not appear in the particle flux. 
We can then obtain the vorticity flux by subtracting the PV flux from the particle flux, i.e., 
$\langle \tilde{v}_x \tilde{\rho} \rangle 
= \langle \tilde{v}_x \tilde{n} \rangle 
- \langle \tilde{v}_x \tilde{q} \rangle$, 
which is
$ \langle \tilde{v}_x \tilde{\rho} \rangle
	= - (D_{n,\text{turb}} - D_q^{\text{res}}) \partial_x \langle n \rangle
	- D_q^{\text{res}} \partial_x \langle \rho \rangle$.
Here, the last term is the flux induced by resonant diffusion. 
The non-diffusive component forms a \textit{residual vorticity flux}, i.e., 
$\Gamma_{\rho}^{Res} = - (D_{n,\text{turb}} - D_q^{\text{res}}) \partial_x \langle n \rangle$.
$\Gamma_{\rho}^{Res}$ is driven by DW turbulence, so it is proportional to the density gradient.
As discussed in Ref. \cite{Diamond_PPCF_2005}, $\Gamma_{\rho}^{Res}$ drives ZF against diffusive mixing and 
the gradient of $\Gamma_{\rho}^{Res}$ can accelerate ZF from rest. 
Note that this mean field calculation of the vorticity flux is applicable to the stationary state, while modulational instability analysis applies only to the stage of ZF growth. 

We then arrive at the DW--ZF system including resonant PV mixing, which is
\begin{equation} \label{eq:model_density}
\frac{\partial}{\partial t} \langle n \rangle = 
\frac{\partial}{\partial x} D_{n,\text{turb}} \frac{\partial}{\partial x} \langle n \rangle
+  \frac{\partial}{\partial x} D_{NL}  \frac{\partial}{\partial x} \langle n \rangle
+ D_{c} \nabla^2 \langle n \rangle,
\end{equation}
\begin{equation} \label{eq:model_vorticity}
\frac{\partial}{\partial t} \langle \rho \rangle = 
\frac{\partial}{\partial x} \left[
(D_{n,\text{turb}} - D_q^{\text{res}}) \frac{\partial}{\partial x} \langle n \rangle
+ D_q^{\text{res}} 
\frac{\partial}{\partial x} \langle \rho \rangle
\right]
- \mu_{c} \langle \rho \rangle
+  \frac{\partial}{\partial x} D_{NL}  \frac{\partial}{\partial x} \langle \rho \rangle
+ \chi_{c} \nabla^2 \langle \rho \rangle,
\end{equation}
\begin{equation} \label{eq:model_pe}
\frac{\partial}{\partial t} \Omega
= D_\Omega \frac{\partial^2}{\partial x^2} \Omega
+ (D_q^{\text{res}} + D_{NL})
\left[\frac{\partial}{\partial x} 
(\langle n \rangle - \langle \rho \rangle) \right]^2
- \varepsilon_c \Omega^{3/2}
+ \gamma_L \Omega.
\end{equation}
This system consists of the equations for mean-field density (Eq. \eqref{eq:model_density}), zonal vorticity (Eq. \eqref{eq:model_vorticity}), and fluctuation PE (Eq. \eqref{eq:model_pe}). 
Initially produced by linear DW instability, the PE of this system is conserved up to frictional dissipation and nonlinear turbulent saturation, which transfer PE to small scales. The evolution of total PE is given by
\begin{equation} \label{eq:enstrophy}
	\frac{\partial}{\partial t} \int dx
	\left[
	\Omega + \frac{(\langle n \rangle - \langle \rho \rangle)^2}{2}
	\right]
	= \int dx \left[
	\gamma_L \Omega - \varepsilon_c \Omega^{3/2}
	- D_{q,c} |\nabla (\langle n \rangle - \langle \rho \rangle)|^2
	- \mu_{c} \langle \rho \rangle^2
	\right].
\end{equation}
The collisional diffusion of zonal PV (the term with $D_{q,c}$ in Eq. \eqref{eq:enstrophy}) is a sink.
In contrast, the turbulent PV diffusion conserves PE between mean field and fluctuations.

As demonstrated by Ref. \cite{Taylor_1915,Diamond_PoF_1991}, vorticity flux is identical to the Reynolds force, and thus drives the ZF. 
The residual vorticity flux excites the ZF, and thus the resonant diffusion is the only damping for ZF in the frictionless limit--i.e., $\mu_c,\chi_c,D_{NL} \rightarrow 0$.
By multiplying Eq. \eqref{eq:model_vorticity} by $\langle \rho \rangle$, we obtain the net production of mean flow enstrophy in the frictionless limit, which is 
\begin{equation} \label{eq:mean_enstrophy}
	\frac{\partial}{\partial t} \int dx \frac{\langle \rho \rangle^2}{2}
	= \int dx  \left[
	- (D_{n,\text{turb}} - D_q^{\text{res}}) \frac{\partial \langle n \rangle}{\partial x}
	\frac{\partial \langle \rho \rangle}{\partial x}
	- D_q^{\text{res}} 
	\left(\frac{\partial \langle \rho \rangle}{\partial x}\right)^2
	\right].
\end{equation}
Note resonant diffusion of zonal vorticity acts to damp and saturate ZF in the frictionless regime--i.e., its contribution to $\partial_t \int dx \langle \rho \rangle ^2 $ is negative definite.

The zonal vorticity profile is stationary when the net flow production is zero, i.e., $\partial_t \int dx \langle \rho \rangle ^2 = 0$. 
Therefore, in the frictionless regime, the stationary vorticity profile is determined by the balance between residual vorticity flux and the resonant vorticity diffusion (i.e., so $\langle \tilde{v}_x \tilde{\rho} \rangle = 0$) which implies 
\begin{equation} \label{eq:vorticity_profile}
	\langle v_y \rangle'' 
	\sim 
	- \frac{c_s}{\rho_s L_n}
	\left(
	1
	-\frac{1}{ \tau_{ck} k_\parallel^2 D_\parallel}
	\frac{k_y^2 \rho_s^2 + L_m^{-2}\rho_s^2}{1+k_y^2 \rho_s^2 + L_m^{-2}\rho_s^2}
	\right).
\end{equation}

In the relevant limit of near-adiabatic electrons, we have $\tau_{ck} k_\parallel^2 D_\parallel \gg 1$. 
The flow scale is then $L_{ZF} \sim (\langle v_y \rangle / c_s)^{1/2} \sqrt{\rho_s L_n}$.
The flow magnitude is obtained using mixing length estimation for the turbulence energy and the fraction $f$: 
$ \langle v_y \rangle^2/c_s^2 \sim f l_{mix}^2/L_n^2$,
where $0<f<1$ is the fraction of turbulence energy coupled to ZF.
The appropriate ``base state'' mixing scale ($l_0$), absent ZF, is the size of an extended cell absent shear, i.e., $l_0 \sim L_n$.
$l_0$ is reduced by ZF shearing, according to the model
$l_{mix}^2 \sim l_0^2 /[1+(\langle v_y \rangle' \tau_c)^2]$. 
For weak or modest ZF shear, the decorrelation time is the eddy turnover time, i.e., $\tau_c \sim \langle \tilde{\rho}^2 \rangle^{-1/2}
\sim l_{mix}/\langle \tilde{v}^2 \rangle^{1/2}$.
The mixing model yields $\langle \tilde{v}^2 \rangle/c_s^2 \sim (1-f) l_{mix}^2/L_n^2$.
Thus, ZF scale is $L_{ZF} \sim f^{1/6}(1-f)^{1/6} \rho_s^{2/3} l_0^{1/3}$
and ZF shear is $|\langle v_y \rangle '| \sim f^{1/6}(1-f)^{1/6} (c_s/L_n) ( l_0 / \rho_s ) ^{1/3}$.
For strong ZF shear, i.e., $\langle v_y \rangle ' \gg$ eddy turnover rate, the interaction of shearing and radial scattering sets the decorrelation time, which yields 
$\tau_c \sim (\langle v_y \rangle'^2 D_q^{\text{res}} l_{mix}^{-2})^{-1/3}$\cite{Biglari_PoF_1990}. 
Due to the strong ZF shear, the radial scattering is resonant, so 
$D_q^{\text{res}} \sim \sum_k |\tilde{v}_x|^2 \delta (\omega_k - k_y \langle v_y \rangle)$
where $\delta (\omega_k - k_\theta \langle v_y \rangle) \sim |\langle v_y \rangle'|^{-1}$.
Hence, the mixing model gives $D_q^{\text{res}} \sim (1-f) (c_s^2/|\langle v_y \rangle'| )(l_{mix}^2/L_n^2)$.
As a result, the ZF scale is $L_{ZF}\sim f^{3/16} (1-f)^{1/8} \rho_s^{5/8} l_0^{3/8}$
and the ZF shear is $|\langle v_y \rangle '| \sim f^{3/16}(1-f)^{1/8} (c_s/L_n) ( l_0/\rho_s ) ^{3/8}$.
In either case, $L_{ZF}$ depends weakly on $f$ and $1-f$ and so is mesoscopic, with the microscale ($\rho_s$) weighed somewhat more strongly than the macroscale ($l_0$). 
Thus, while $L_{ZF}$ is a hybrid of $\rho_s$ and $l_0$, it tends slightly toward the microscale.
Observe a mesoscopic ZF scale is frequently assumed (i.e., $L_{ZF}\sim\sqrt{\rho_s L_n}$ is a standard guesstimate), here it is derived from the analysis.
The ZF shears in both cases are quite similar and robust. 
Hence, the case of strong ZF shear--and thus flow resonance--is more likely relevant to the frictionless DW--ZF system discussed here. 

When $\tau_{ck} k_\parallel^2 D_\parallel$ is comparable to unity, $L_{ZF}$ is linked to the mode scale. In that case, the resonance between DW and ZF regulates the flow structure by modifying the local mode scale. 
In the hydrodynamic limit (i.e., $\tau_{ck} k_\parallel^2 D_\parallel \ll 1$), the generation and saturation of ZF need to be reconsidered, because the DW model discussed here is not directly applicable.
Overall, the mesoscopic ZF appears as a limiting case with near-adiabatic electrons (i.e., $\tau_{ck} k_\parallel^2 D_\parallel \gg 1$). 

Frictionless saturation induced by resonant vorticity mixing can be elucidated using a 0D predator--prey model of the DW--ZF system. 
Ignoring the evolution of $\langle n \rangle$, the total mean-field PE is related to the zonal vorticity through 
$V''^2 \sim \int dx \langle v_y \rangle'^2/L_{ZF}^2 \sim \int dx \langle \rho \rangle ^2/L_{ZF}^2 $.
The total fluctuation PE is
$E \equiv \int dx \Omega$.
The \textit{net} mean-field PE is produced by 
$\langle \tilde{v}_x \tilde{\rho} \rangle V'' =  \Gamma_{\rho}^{Res} V'' - D_q^{\text{res}} V''^2 \sim \alpha_1 E |V''|- \alpha_2 V''^2 E$.
Therefore, with both frictional damping and nonlinear damping by tertiary instability included, the 0D predator--prey system is 
\begin{equation} \label{eq:predator}
\frac{L_{ZF}^2}{2} \frac{d V''^2}{dt}
= \alpha_1 |V''| E
- \alpha_2 V''^2 E
- D_{NL} V''^2
- \mu_{c} V''^2,
\end{equation}
\begin{equation} \label{eq:prey}
\frac{d E}{dt}
= 
- \alpha_1 |V''| E
+ \alpha_2 V''^2 E
+ D_{NL} V''^2
- \varepsilon_c E^{3/2}
+ \gamma_L E.
\end{equation}
Here, baseline (i.e., without flow) nonlinear saturation of turbulence is through the forward cascade of PE.
Ultimately, PE is dissipated by collisional diffusion at small scales.
The linear growth of energy is due to the (linear) instability of fluctuations. 

This new predator--prey model conserves PE
and includes resonant PV mixing.
Note that the quantities here have been integrated over space, and so should be interpreted as characteristic magnitudes.
Though this simplified 0D model is semi-quantitative, we can use it to obtain useful insights. 

The flow and turbulence states are set by fixed points of the system and are summarized in Table \ref{table:states}. We ignore the nonlinear flow damping by tertiary instability as irrelevant to the Dimits up-shift regime. Therefore, we obtain the zonal vorticity from Eq. \eqref{eq:predator}, which is
$|V''| = \alpha_1 E/(\alpha_2 E + \mu_c)$.
We next discuss three regimes---the frictionless regime, the weakly frictional regime, and the strongly frictional regime. In particular, we emphasize \textit{what determines the turbulence level} and \textit{what affects the flow} in quasi-marginal turbulence. 

\begin{table}
	\centering
	\setlength{\tabcolsep}{5pt}
	\renewcommand{\arraystretch}{1.2}
	\caption{Stationary flow and turbulence level are compared among regimes with different frictional drags.
    \label{table:states}}
	\begin{tabular}{c c c c}
		\hline \hline
		Regime & Frictionless & Weakly Frictional & Strongly Frictional \\
		\hline 
		$\mu_c$ & $\mu_c \ll \alpha_2 E$ & $\alpha_2 E \ll \mu_c \ll 4 \gamma_L \alpha_1^2 /\varepsilon_c^2$ & $\mu_c \gg 4 \gamma_L \alpha_1^2 /\varepsilon_c^2$ \\
		$|V''|$ & $\frac{\alpha_1}{\alpha_2}$ & $\frac{\alpha_1 \gamma_L^2}{\mu_c \varepsilon_c^2}$ & $\frac{\gamma_L}{\alpha_1}$ \\ [6pt]
		$E$ & $\frac{\gamma_L^2}{\varepsilon_c^2}$ & $\frac{\gamma_L^2}{\varepsilon_c^2}$ & $\frac{\gamma_L \mu_c}{\alpha_1^2}$ \\ [6pt]
		\hline \hline
	\end{tabular}
\end{table}

\textit{Frictionless regime:} 
In the frictionless regime, flow drag is negligible compared to vorticity diffusion, i.e. 
$\mu_c \ll \alpha_2 E$.
Turbulence energy is independent of the flow state, because $E$ is determined only by the balance between linear instability drive ($\gamma_L$) and nonlinear dissipation of PE due to forward cascade ($\varepsilon_c \Omega^{1/2} \sim \varepsilon_c E^{1/2}$).
The turbulence state is then set by $\gamma_L \sim \varepsilon_c E^{1/2}$, yielding $E \sim (\gamma_L/\varepsilon_c)^2$.
When the linear drive is weak, i.e. $\gamma_L/\varepsilon_c < 1$, the turbulence becomes marginal, with $E\ll1$.
The balance between residual vorticity flux and resonant vorticity diffusion sets the zonal vorticity. In this balance, the turbulence intensity cancels to leading order. This means \textit{there can be significant ZF, even when the turbulence is weak}. Therefore, this new frictionless saturation mechanism, induced by resonant PV mixing, is effective for turbulence near marginality.

\textit{Weakly frictional regime:}
When the frictional drag exceeds turbulent diffusion, i.e. 
$\mu_c \gg \alpha_2 E$,
zonal vorticity is linked to the turbulence strength, yielding
$|V''| = \alpha_1 E/\mu_c$.
This follows because the flow is driven by turbulence, and collisions are the major source of flow damping.
The weakly frictional regime is a hybrid of the frictionless and strongly frictional regimes. On one hand, the turbulence level is independent of flow damping (i.e., $E \sim (\gamma_L/\varepsilon_c)^2$), as for the frictionless regime. 
On the other hand, the flow depends on the turbulence level (i.e., $|V''| \sim E \sim (\gamma_L/\varepsilon_c)^2$), meaning that when the turbulence is very weak, the flow is also weak. This is because the turbulence needs to be strong enough to overcome frictional damping in order to drive a significant ZF. 

\textit{Strongly frictional regime:}
When the frictional flow damping is strong, i.e., $\mu_c \gg 4 \gamma_L \alpha_1^2 /\varepsilon_c^2$, the turbulence energy is set by the flow damping,
recovering the scaling trends of previous predator--prey models, i.e., $E\sim \mu_c$ and $|V''| \sim \gamma_L$.

Note that $|V''|$ considers only zonal flows generated by drift wave turbulence. Therefore, without drift waves, both zonal flow generation and saturation go to zero, and thus $|V''| = 0$. 
A truly laminar state is possible only when externally driven flows are given.

The new predator--prey model presented here does not depend sensitively on the specific turbulence type or features. To compare with the results calculated from the zonal vorticity equation, we use DW instability as an example. The coefficients are
\begin{equation}
	\alpha_1 = \frac{k_y^2\rho_s c_s}{L_n} \left( \tau_{ck} - \frac{1}{k_\parallel^2 D_\parallel}
	\frac{k_y^2 \rho_s^2 + L_m^{-2}\rho_s^2}{1+k_y^2 \rho_s^2 + L_m^{-2}\rho_s^2} \right),
\end{equation}
\begin{equation}
	\alpha_2 = k_y^2\rho_s^2 \tau_{ck}.
\end{equation}
As a result, in the frictionless regime, the stationary state zonal vorticity gradient emerges as
\begin{equation}
	|V''| = \frac{\alpha_1}{\alpha_2}
	= \frac{c_s}{\rho_s L_n}
	\left(
	1
	-\frac{1}{ \tau_{ck} k_\parallel^2 D_\parallel}
	\frac{k_y^2 \rho_s^2 + L_m^{-2}\rho_s^2}{1+k_y^2 \rho_s^2 + L_m^{-2}\rho_s^2}
	\right),
\end{equation}
which is consistent with Eq. \eqref{eq:vorticity_profile}.
The vorticity gradient measures the jump across the flow shear field.
Thus, the ZF profile can be deduced from the zonal vorticity by specifying boundary conditions.
For ZF, vorticity is identical to shear, which is of greater interest than the flow velocity. 

In summary, the resonant scattering of vorticity tends to dissipate zonal flows. 
In this paper, we have developed this new mechanism explicitly for a model of electron drift wave turbulence. 
This dissipation mechanism of zonal flows is generic to electron drift wave turbulence.
It follows from the nonlinear structure of the vorticity equation (i.e., $\nabla \cdot \mathbf{J} = 0$, where $\mathbf{J}$ is the plasma current), which is generic to \textit{all} drift wave models.
The question of how strong this effect is compared to other dissipation effects, i.e., the relative importance of different mechanisms, is beyond the scope of this work. 
In magnetic fusion plasmas, the robustness of tertiary modes is dubious. 
The tertiary instability observed in simulations \cite{Rogers_PRL_2000} requires an artificial increase of the $E\times B$ flow shear. 
Therefore, resonant vorticity mixing is a generic, broadly active and viable saturation mechanism for zonal flows. 
In other systems, such as where magnetic shear is weak, both tertiary modes and the resonant vorticity mixing may have significant effects, but the relative importance of these two mechanisms needs further investigation. 
This is left for future works.

While ZF scale is often \textit{assumed}, the new model discussed here \textit{calculates} the saturated flow scale and flow shear in the frictionless limit. In the limiting case with near-adiabatic electrons (i.e., $\tau_{ck} k_\parallel^2 D_\parallel \gg 1$),
the ZF scale is seen to be mesoscopic, i.e., 
$L_{ZF}\sim f^{3/16} (1-f)^{1/8} \rho_s^{5/8} l_0^{3/8}$. 
$f$ can be determined by simulation and experimental measurements. 
The dependence of zonal flow scale and shear on $f$ is weak because of the small exponent in the results.

The model discussed here addresses the long-standing question of ``how close is `close'" in near-marginal systems.
It is effective in both near-marginal turbulence and in the frictionless regime.
Thus, when expanded to \textit{1D}, it can be used to study avalanches and staircase formation\cite{Dif_PRE_2010,Newman_PoP_1996}.
In 1D, avalanching induces variability of profiles, and thus of local growth rates. 
The scaling $E \sim \gamma_L^2$ suggests a variability-dominated state can result when $\gamma_L \rightarrow 0$.
This follows because $\gamma_L$ has an exponent $>1$, which holds true as long as the self-saturation of fluctuation PE exhibits the dependence $\varepsilon_c \Omega^{1+p}$, where $0<p<1$.
Thus, the scaling of $E$ with $\gamma_L$ is stronger than the conventional weak turbulence result.
The local linear growth rate is then set by both equilibrium (mean) and variable (i.e., avalanche-induced) profile gradients, i.e., $\gamma_L = \overline{\gamma}_L + \tilde{\gamma}_L$.
As a result of resonant PV mixing in the frictionless regime, the turbulence state is determined by $E \sim \gamma_L^2 \sim \overline{\gamma}_L^2 + \tilde{\gamma}_L^2$.
$\overline{\gamma}_L$ is determined by the difference between mean profile gradient and critical gradient.
In near-marginal turbulence, the mean gradient approaches the critical gradient, so $\overline{\gamma}_L \rightarrow 0$. 
Thus, there the turbulence state is primarily controlled by noise from avalanche variability, i.e., $E \sim \tilde{\gamma}_L^2 \gg \overline{\gamma}_L^2$. 
Such noise is produced by avalanching, which stochastically modulates the driving gradient.
In this case, the predator--prey model must be treated as a set of coupled stochastic differential equations.
In 1D, the relevant system is a nonlinear reaction--diffusion model like that of Eq. \eqref{eq:model_vorticity} and \eqref{eq:model_pe}, including multiplicative noise. 
The results in this work thus define the boundary for ``marginality".
The turbulence energy scales with the dimensionless ratio $(\gamma_L/\varepsilon_c)^2$, where $\varepsilon_c$ is the dissipation rate of PE.
Therefore, the turbulence can be ``marginal" when the equilibrium growth rate $\overline{\gamma}_L < \varepsilon_c$.
This gives a basis with which to define the extent of the ``near-marginal regime''.
The theory presented here predicts a smooth transition from the near-marginal regime to the strong turbulence regime, as the free energy source (e.g., $\nabla n_0$ and/or $\nabla T$) increases, i.e., $E \sim (\gamma_L/\varepsilon_c)^2$.

The Dimits up-shift regime spans low to zero collisionality and consists of weak turbulence near marginality.
ZF saturation induced by resonant PV mixing is effective in both the frictionless regime and for near-marginal turbulence, and thus is compatible with the physics of the Dimits up-shift regime.
Resonance regulates ZF saturation in the frictionless regime, eliminating the need to invoke or provoke tertiary instability. 
The saturated flow does \textit{not} depend on the turbulence intensity. Hence, there can be significant ZF for near-marginal turbulence, absent frictional damping.
In the upshift regime, turbulence is near-marginal, rather than zero, i.e., $0<E \ll 1$ and $\overline{\gamma}_L > 0$ in the upshift regime. 
The model predicts that the boundary of this near-marginal regime is determined by $\overline{\gamma}_L < \epsilon_c$, where $\epsilon_c$ is the nonlinear dissipation rate of potential enstrophy.
This means the nonlinear critical gradient is determined by $\epsilon_c$.
Note that resonant PV mixing is essentially different from the zonal flow saturation by tertiary modes.
Their relative importance can be accounted by comparing the transition from near-marginal state to strong turbulence state \cite{St_JPP_2017} predicted by these two zonal flow saturation mechanisms.

\begin{acknowledgments}
	The authors are grateful to Z. B. Guo, A. Ashourvan, and H. Che for inspiring and insightful discussions. 
	We acknowledge useful and interesting discussions at the Festival de Th{\'e}orie.
	This work was supported by the U.S. Department of Energy, Office of Science, OFES, under Award Number DE-FG02-04ER54738.
\end{acknowledgments}


\begin{thebibliography}{31}
	\expandafter\ifx\csname natexlab\endcsname\relax\def\natexlab#1{#1}\fi
	\expandafter\ifx\csname bibnamefont\endcsname\relax
	\def\bibnamefont#1{#1}\fi
	\expandafter\ifx\csname bibfnamefont\endcsname\relax
	\def\bibfnamefont#1{#1}\fi
	\expandafter\ifx\csname citenamefont\endcsname\relax
	\def\citenamefont#1{#1}\fi
	\expandafter\ifx\csname url\endcsname\relax
	\def\url#1{\texttt{#1}}\fi
	\expandafter\ifx\csname urlprefix\endcsname\relax\def\urlprefix{URL }\fi
	\providecommand{\bibinfo}[2]{#2}
	\providecommand{\eprint}[2][]{\url{#2}}
	
	\bibitem[{\citenamefont{Moody}(1944)}]{Moody_1944}
	\bibinfo{author}{\bibfnamefont{L.~F.} \bibnamefont{Moody}},
	\bibinfo{journal}{Trans. of the ASME} \textbf{\bibinfo{volume}{66}},
	\bibinfo{pages}{671} (\bibinfo{year}{1944}).
	
	\bibitem[{\citenamefont{Berg\'{e} and Dubois}(1984)}]{Berge_1984}
	\bibinfo{author}{\bibfnamefont{P.}~\bibnamefont{Berg\'{e}}} \bibnamefont{and}
	\bibinfo{author}{\bibfnamefont{M.}~\bibnamefont{Dubois}},
	\bibinfo{journal}{Contemporary Physics} \textbf{\bibinfo{volume}{25}},
	\bibinfo{pages}{535} (\bibinfo{year}{1984}).
	
	\bibitem[{\citenamefont{Garbet et~al.}(2004)\citenamefont{Garbet, Mantica,
			Angioni, Asp, Baranov, Bourdelle, Budny, Crisanti, Cordey, Garzotti
			et~al.}}]{Garbet_PPCF_2004}
	\bibinfo{author}{\bibfnamefont{X.}~\bibnamefont{Garbet}},
	\bibinfo{author}{\bibfnamefont{P.}~\bibnamefont{Mantica}},
	\bibinfo{author}{\bibfnamefont{C.}~\bibnamefont{Angioni}},
	\bibinfo{author}{\bibfnamefont{E.}~\bibnamefont{Asp}},
	\bibinfo{author}{\bibfnamefont{Y.}~\bibnamefont{Baranov}},
	\bibinfo{author}{\bibfnamefont{C.}~\bibnamefont{Bourdelle}},
	\bibinfo{author}{\bibfnamefont{R.}~\bibnamefont{Budny}},
	\bibinfo{author}{\bibfnamefont{F.}~\bibnamefont{Crisanti}},
	\bibinfo{author}{\bibfnamefont{G.}~\bibnamefont{Cordey}},
	\bibinfo{author}{\bibfnamefont{L.}~\bibnamefont{Garzotti}},
	\bibnamefont{et~al.}, \bibinfo{journal}{Plasma Physics and Controlled Fusion}
	\textbf{\bibinfo{volume}{46}}, \bibinfo{pages}{B557} (\bibinfo{year}{2004}).
	
	\bibitem[{\citenamefont{McKee et~al.}(2001)\citenamefont{McKee, Petty, Waltz,
			Fenzi, Fonck, Kinsey, Luce, Burrell, Baker, Doyle et~al.}}]{McKee_NF_2001}
	\bibinfo{author}{\bibfnamefont{G.}~\bibnamefont{McKee}},
	\bibinfo{author}{\bibfnamefont{C.}~\bibnamefont{Petty}},
	\bibinfo{author}{\bibfnamefont{R.}~\bibnamefont{Waltz}},
	\bibinfo{author}{\bibfnamefont{C.}~\bibnamefont{Fenzi}},
	\bibinfo{author}{\bibfnamefont{R.}~\bibnamefont{Fonck}},
	\bibinfo{author}{\bibfnamefont{J.}~\bibnamefont{Kinsey}},
	\bibinfo{author}{\bibfnamefont{T.}~\bibnamefont{Luce}},
	\bibinfo{author}{\bibfnamefont{K.}~\bibnamefont{Burrell}},
	\bibinfo{author}{\bibfnamefont{D.}~\bibnamefont{Baker}},
	\bibinfo{author}{\bibfnamefont{E.}~\bibnamefont{Doyle}},
	\bibnamefont{et~al.}, \bibinfo{journal}{Nuclear Fusion}
	\textbf{\bibinfo{volume}{41}}, \bibinfo{pages}{1235} (\bibinfo{year}{2001}).
	
	\bibitem[{\citenamefont{Charney}(1948)}]{Charney_1948}
	\bibinfo{author}{\bibfnamefont{J.~G.} \bibnamefont{Charney}},
	\bibinfo{journal}{Geofysiske Publikasjoner} \textbf{\bibinfo{volume}{17}}
	(\bibinfo{year}{1948}).
	
	\bibitem[{\citenamefont{Diamond et~al.}(2005)\citenamefont{Diamond, Itoh, Itoh,
			and Hahm}}]{Diamond_PPCF_2005}
	\bibinfo{author}{\bibfnamefont{P.~H.} \bibnamefont{Diamond}},
	\bibinfo{author}{\bibfnamefont{S.~I.} \bibnamefont{Itoh}},
	\bibinfo{author}{\bibfnamefont{K.}~\bibnamefont{Itoh}}, \bibnamefont{and}
	\bibinfo{author}{\bibfnamefont{T.~S.} \bibnamefont{Hahm}},
	\bibinfo{journal}{Plasma Physics and Controlled Fusion}
	\textbf{\bibinfo{volume}{47}}, \bibinfo{pages}{R35} (\bibinfo{year}{2005}).
	
	\bibitem[{\citenamefont{Diamond et~al.}(1994)\citenamefont{Diamond, Liang,
			Carreras, and Terry}}]{Diamond_PRL_1994}
	\bibinfo{author}{\bibfnamefont{P.~H.} \bibnamefont{Diamond}},
	\bibinfo{author}{\bibfnamefont{Y.-M.} \bibnamefont{Liang}},
	\bibinfo{author}{\bibfnamefont{B.~A.} \bibnamefont{Carreras}},
	\bibnamefont{and} \bibinfo{author}{\bibfnamefont{P.~W.} \bibnamefont{Terry}},
	\bibinfo{journal}{Phys. Rev. Lett.} \textbf{\bibinfo{volume}{72}},
	\bibinfo{pages}{2565} (\bibinfo{year}{1994}).
	
	\bibitem[{\citenamefont{Kobayashi et~al.}(2015)\citenamefont{Kobayashi,
			G{\"u}rcan, and Diamond}}]{Kobayashi_PoP_2015}
	\bibinfo{author}{\bibfnamefont{S.}~\bibnamefont{Kobayashi}},
	\bibinfo{author}{\bibfnamefont{{\"O}.~D.} \bibnamefont{G{\"u}rcan}},
	\bibnamefont{and} \bibinfo{author}{\bibfnamefont{P.~H.}
		\bibnamefont{Diamond}}, \bibinfo{journal}{Physics of Plasmas}
	\textbf{\bibinfo{volume}{22}}, \bibinfo{pages}{090702}
	(\bibinfo{year}{2015}).
	
	\bibitem[{\citenamefont{G{\"u}rcan and Diamond}(2015)}]{Gurcan_ZF_2015}
	\bibinfo{author}{\bibfnamefont{{\"O}.~D.} \bibnamefont{G{\"u}rcan}}
	\bibnamefont{and} \bibinfo{author}{\bibfnamefont{P.~H.}
		\bibnamefont{Diamond}}, \bibinfo{journal}{Journal of Physics A: Mathematical
		and Theoretical} \textbf{\bibinfo{volume}{48}}, \bibinfo{pages}{293001}
	(\bibinfo{year}{2015}).
	
	\bibitem[{\citenamefont{Staebler}(2004)}]{Staebler_PoP_2004}
	\bibinfo{author}{\bibfnamefont{G.~M.} \bibnamefont{Staebler}},
	\bibinfo{journal}{Physics of Plasmas} \textbf{\bibinfo{volume}{11}},
	\bibinfo{pages}{1064} (\bibinfo{year}{2004}).
	
	\bibitem[{\citenamefont{Dimits et~al.}(1996)\citenamefont{Dimits, Williams,
			Byers, and Cohen}}]{Dimits_PRL_1996}
	\bibinfo{author}{\bibfnamefont{A.~M.} \bibnamefont{Dimits}},
	\bibinfo{author}{\bibfnamefont{T.~J.} \bibnamefont{Williams}},
	\bibinfo{author}{\bibfnamefont{J.~A.} \bibnamefont{Byers}}, \bibnamefont{and}
	\bibinfo{author}{\bibfnamefont{B.~I.} \bibnamefont{Cohen}},
	\bibinfo{journal}{Phys. Rev. Lett.} \textbf{\bibinfo{volume}{77}},
	\bibinfo{pages}{71} (\bibinfo{year}{1996}).
	
	\bibitem[{\citenamefont{Ernst et~al.}(2009)\citenamefont{Ernst, Lang, Nevins,
			Hoffman, Chen, Dorland, and Parker}}]{Ernst_PoP_2009}
	\bibinfo{author}{\bibfnamefont{D.}~\bibnamefont{Ernst}},
	\bibinfo{author}{\bibfnamefont{J.}~\bibnamefont{Lang}},
	\bibinfo{author}{\bibfnamefont{W.}~\bibnamefont{Nevins}},
	\bibinfo{author}{\bibfnamefont{M.}~\bibnamefont{Hoffman}},
	\bibinfo{author}{\bibfnamefont{Y.}~\bibnamefont{Chen}},
	\bibinfo{author}{\bibfnamefont{W.}~\bibnamefont{Dorland}}, \bibnamefont{and}
	\bibinfo{author}{\bibfnamefont{S.}~\bibnamefont{Parker}},
	\bibinfo{journal}{Physics of Plasmas} \textbf{\bibinfo{volume}{16}},
	\bibinfo{pages}{055906} (\bibinfo{year}{2009}).
	
	\bibitem[{\citenamefont{Lang et~al.}(2008)\citenamefont{Lang, Parker, and
			Chen}}]{Lang_PoP_2008}
	\bibinfo{author}{\bibfnamefont{J.}~\bibnamefont{Lang}},
	\bibinfo{author}{\bibfnamefont{S.~E.} \bibnamefont{Parker}},
	\bibnamefont{and} \bibinfo{author}{\bibfnamefont{Y.}~\bibnamefont{Chen}},
	\bibinfo{journal}{Physics of Plasmas} \textbf{\bibinfo{volume}{15}},
	\bibinfo{pages}{055907} (\bibinfo{year}{2008}).
	
	\bibitem[{\citenamefont{Wang et~al.}(2006)\citenamefont{Wang, Lin, Tang, Lee,
			Ethier, Lewandowski, Rewoldt, Hahm, and Manickam}}]{Wang_PoP_2006}
	\bibinfo{author}{\bibfnamefont{W.}~\bibnamefont{Wang}},
	\bibinfo{author}{\bibfnamefont{Z.}~\bibnamefont{Lin}},
	\bibinfo{author}{\bibfnamefont{W.}~\bibnamefont{Tang}},
	\bibinfo{author}{\bibfnamefont{W.}~\bibnamefont{Lee}},
	\bibinfo{author}{\bibfnamefont{S.}~\bibnamefont{Ethier}},
	\bibinfo{author}{\bibfnamefont{J.}~\bibnamefont{Lewandowski}},
	\bibinfo{author}{\bibfnamefont{G.}~\bibnamefont{Rewoldt}},
	\bibinfo{author}{\bibfnamefont{T.}~\bibnamefont{Hahm}}, \bibnamefont{and}
	\bibinfo{author}{\bibfnamefont{J.}~\bibnamefont{Manickam}},
	\bibinfo{journal}{Physics of Plasmas} \textbf{\bibinfo{volume}{13}},
	\bibinfo{pages}{092505} (\bibinfo{year}{2006}).
	
	\bibitem[{\citenamefont{Jenko and Kendl}(2002)}]{Jenko_PoP_2002}
	\bibinfo{author}{\bibfnamefont{F.}~\bibnamefont{Jenko}} \bibnamefont{and}
	\bibinfo{author}{\bibfnamefont{A.}~\bibnamefont{Kendl}},
	\bibinfo{journal}{Physics of Plasmas} \textbf{\bibinfo{volume}{9}},
	\bibinfo{pages}{4103} (\bibinfo{year}{2002}).
	
	\bibitem[{\citenamefont{Rogers et~al.}(2000)\citenamefont{Rogers, Dorland, and
			Kotschenreuther}}]{Rogers_PRL_2000}
	\bibinfo{author}{\bibfnamefont{B.~N.} \bibnamefont{Rogers}},
	\bibinfo{author}{\bibfnamefont{W.}~\bibnamefont{Dorland}}, \bibnamefont{and}
	\bibinfo{author}{\bibfnamefont{M.}~\bibnamefont{Kotschenreuther}},
	\bibinfo{journal}{Phys. Rev. Lett.} \textbf{\bibinfo{volume}{85}},
	\bibinfo{pages}{5336} (\bibinfo{year}{2000}).
	
	\bibitem[{\citenamefont{Kim and Diamond}(2002)}]{Kim_PoP_2002}
	\bibinfo{author}{\bibfnamefont{E.-J.} \bibnamefont{Kim}} \bibnamefont{and}
	\bibinfo{author}{\bibfnamefont{P.~H.} \bibnamefont{Diamond}},
	\bibinfo{journal}{Physics of Plasmas} \textbf{\bibinfo{volume}{9}},
	\bibinfo{pages}{4530} (\bibinfo{year}{2002}).
	
	\bibitem[{\citenamefont{Mantica et~al.}(2011)\citenamefont{Mantica, Angioni,
			Challis, Colyer, Frassinetti, Hawkes, Johnson, Tsalas, deVries, Weiland
			et~al.}}]{Mantica_PRL_2011}
	\bibinfo{author}{\bibfnamefont{P.}~\bibnamefont{Mantica}},
	\bibinfo{author}{\bibfnamefont{C.}~\bibnamefont{Angioni}},
	\bibinfo{author}{\bibfnamefont{C.}~\bibnamefont{Challis}},
	\bibinfo{author}{\bibfnamefont{G.}~\bibnamefont{Colyer}},
	\bibinfo{author}{\bibfnamefont{L.}~\bibnamefont{Frassinetti}},
	\bibinfo{author}{\bibfnamefont{N.}~\bibnamefont{Hawkes}},
	\bibinfo{author}{\bibfnamefont{T.}~\bibnamefont{Johnson}},
	\bibinfo{author}{\bibfnamefont{M.}~\bibnamefont{Tsalas}},
	\bibinfo{author}{\bibfnamefont{P.~C.} \bibnamefont{deVries}},
	\bibinfo{author}{\bibfnamefont{J.}~\bibnamefont{Weiland}},
	\bibnamefont{et~al.}, \bibinfo{journal}{Phys. Rev. Lett.}
	\textbf{\bibinfo{volume}{107}}, \bibinfo{pages}{135004}
	(\bibinfo{year}{2011}).
	
	\bibitem[{\citenamefont{Chiueh et~al.}(1986)\citenamefont{Chiueh, Terry,
			Diamond, and Sedlak}}]{Chiueh_PoF_1986}
	\bibinfo{author}{\bibfnamefont{T.}~\bibnamefont{Chiueh}},
	\bibinfo{author}{\bibfnamefont{P.}~\bibnamefont{Terry}},
	\bibinfo{author}{\bibfnamefont{P.}~\bibnamefont{Diamond}}, \bibnamefont{and}
	\bibinfo{author}{\bibfnamefont{J.}~\bibnamefont{Sedlak}},
	\bibinfo{journal}{The Physics of fluids} \textbf{\bibinfo{volume}{29}},
	\bibinfo{pages}{231} (\bibinfo{year}{1986}).
	
	\bibitem[{\citenamefont{Galeev et~al.}(1977)\citenamefont{Galeev, Sagdeev,
			Shapiro, and Shevchenko}}]{Galeev_1977}
	\bibinfo{author}{\bibfnamefont{A.}~\bibnamefont{Galeev}},
	\bibinfo{author}{\bibfnamefont{R.}~\bibnamefont{Sagdeev}},
	\bibinfo{author}{\bibfnamefont{V.}~\bibnamefont{Shapiro}}, \bibnamefont{and}
	\bibinfo{author}{\bibfnamefont{V.}~\bibnamefont{Shevchenko}},
	\bibinfo{journal}{Zhurnal Eksperimental'noi i Teoreticheskoi Fiziki}
	\textbf{\bibinfo{volume}{73}}, \bibinfo{pages}{1352} (\bibinfo{year}{1977}).
	
	\bibitem[{\citenamefont{Che et~al.}(2017)\citenamefont{Che, Goldstein, Diamond,
			and Sagdeev}}]{Che_PNAS_2017}
	\bibinfo{author}{\bibfnamefont{H.}~\bibnamefont{Che}},
	\bibinfo{author}{\bibfnamefont{M.~L.} \bibnamefont{Goldstein}},
	\bibinfo{author}{\bibfnamefont{P.~H.} \bibnamefont{Diamond}},
	\bibnamefont{and} \bibinfo{author}{\bibfnamefont{R.~Z.}
		\bibnamefont{Sagdeev}}, \bibinfo{journal}{Proceedings of the National Academy
		of Sciences} \textbf{\bibinfo{volume}{114}}, \bibinfo{pages}{1502}
	(\bibinfo{year}{2017}).
	
	\bibitem[{sup()}]{supplement}
	\bibinfo{note}{See Supplemental Material at [URL will be inserted by publisher]
		for comparison between resonant vorticity mixing and Landau damping.}
	
	\bibitem[{\citenamefont{Hasegawa and Wakatani}(1983)}]{Hasegawa_PRL_1983}
	\bibinfo{author}{\bibfnamefont{A.}~\bibnamefont{Hasegawa}} \bibnamefont{and}
	\bibinfo{author}{\bibfnamefont{M.}~\bibnamefont{Wakatani}},
	\bibinfo{journal}{Physical Review Letters} \textbf{\bibinfo{volume}{50}},
	\bibinfo{pages}{682} (\bibinfo{year}{1983}).
	
	\bibitem[{\citenamefont{Ashourvan and Diamond}(2016)}]{Ashourvan_PRE_2016}
	\bibinfo{author}{\bibfnamefont{A.}~\bibnamefont{Ashourvan}} \bibnamefont{and}
	\bibinfo{author}{\bibfnamefont{P.~H.} \bibnamefont{Diamond}},
	\bibinfo{journal}{Phys. Rev. E} \textbf{\bibinfo{volume}{94}},
	\bibinfo{pages}{051202} (\bibinfo{year}{2016}).
	
	\bibitem[{\citenamefont{Ashourvan and Diamond}(2017)}]{Ashourvan_PoP_2017}
	\bibinfo{author}{\bibfnamefont{A.}~\bibnamefont{Ashourvan}} \bibnamefont{and}
	\bibinfo{author}{\bibfnamefont{P.~H.} \bibnamefont{Diamond}},
	\bibinfo{journal}{Physics of Plasmas} \textbf{\bibinfo{volume}{24}},
	\bibinfo{pages}{012305} (\bibinfo{year}{2017}).
	
	\bibitem[{\citenamefont{Taylor}(1915)}]{Taylor_1915}
	\bibinfo{author}{\bibfnamefont{G.~I.} \bibnamefont{Taylor}},
	\bibinfo{journal}{Philosophical Transactions of the Royal Society of London.
		Series A, Containing Papers of a Mathematical or Physical Character}
	\textbf{\bibinfo{volume}{215}}, \bibinfo{pages}{1} (\bibinfo{year}{1915}).
	
	\bibitem[{\citenamefont{Diamond and Kim}(1991)}]{Diamond_PoF_1991}
	\bibinfo{author}{\bibfnamefont{P.~H.} \bibnamefont{Diamond}} \bibnamefont{and}
	\bibinfo{author}{\bibfnamefont{Y.}~\bibnamefont{Kim}},
	\bibinfo{journal}{Physics of Fluids B: Plasma Physics}
	\textbf{\bibinfo{volume}{3}}, \bibinfo{pages}{1626} (\bibinfo{year}{1991}).
	
	\bibitem[{\citenamefont{Biglari et~al.}(1990)\citenamefont{Biglari, Diamond,
			and Terry}}]{Biglari_PoF_1990}
	\bibinfo{author}{\bibfnamefont{H.}~\bibnamefont{Biglari}},
	\bibinfo{author}{\bibfnamefont{P.~H.} \bibnamefont{Diamond}},
	\bibnamefont{and} \bibinfo{author}{\bibfnamefont{P.~W.} \bibnamefont{Terry}},
	\bibinfo{journal}{Physics of Fluids B: Plasma Physics}
	\textbf{\bibinfo{volume}{2}}, \bibinfo{pages}{1} (\bibinfo{year}{1990}).
	
	\bibitem[{\citenamefont{Dif-Pradalier et~al.}(2010)\citenamefont{Dif-Pradalier,
			Diamond, Grandgirard, Sarazin, Abiteboul, Garbet, Ghendrih, Strugarek, Ku,
			and Chang}}]{Dif_PRE_2010}
	\bibinfo{author}{\bibfnamefont{G.}~\bibnamefont{Dif-Pradalier}},
	\bibinfo{author}{\bibfnamefont{P.~H.} \bibnamefont{Diamond}},
	\bibinfo{author}{\bibfnamefont{V.}~\bibnamefont{Grandgirard}},
	\bibinfo{author}{\bibfnamefont{Y.}~\bibnamefont{Sarazin}},
	\bibinfo{author}{\bibfnamefont{J.}~\bibnamefont{Abiteboul}},
	\bibinfo{author}{\bibfnamefont{X.}~\bibnamefont{Garbet}},
	\bibinfo{author}{\bibfnamefont{P.}~\bibnamefont{Ghendrih}},
	\bibinfo{author}{\bibfnamefont{A.}~\bibnamefont{Strugarek}},
	\bibinfo{author}{\bibfnamefont{S.}~\bibnamefont{Ku}}, \bibnamefont{and}
	\bibinfo{author}{\bibfnamefont{C.~S.} \bibnamefont{Chang}},
	\bibinfo{journal}{Phys. Rev. E} \textbf{\bibinfo{volume}{82}},
	\bibinfo{pages}{025401} (\bibinfo{year}{2010}).
	
	\bibitem[{\citenamefont{Newman et~al.}(1996)\citenamefont{Newman, Carreras,
			Diamond, and Hahm}}]{Newman_PoP_1996}
	\bibinfo{author}{\bibfnamefont{D.~E.} \bibnamefont{Newman}},
	\bibinfo{author}{\bibfnamefont{B.~A.} \bibnamefont{Carreras}},
	\bibinfo{author}{\bibfnamefont{P.~H.} \bibnamefont{Diamond}},
	\bibnamefont{and} \bibinfo{author}{\bibfnamefont{T.~S.} \bibnamefont{Hahm}},
	\bibinfo{journal}{Physics of Plasmas} \textbf{\bibinfo{volume}{3}},
	\bibinfo{pages}{1858} (\bibinfo{year}{1996}).
	
	\bibitem[{\citenamefont{St-Onge}(2017)}]{St_JPP_2017}
	\bibinfo{author}{\bibfnamefont{D.~A.} \bibnamefont{St-Onge}},
	\bibinfo{journal}{Journal of Plasma Physics} \textbf{\bibinfo{volume}{83}}
	(\bibinfo{year}{2017}).
	
\end{thebibliography}


\end{document}